%Paper: astro-ph/9406049
%From: strauss@guinness.ias.edu (Michael Strauss)
%Date: Thu, 16 Jun 94 14:03:18 EDT

%This is standard TeX, and does not include the figures. You can pick
%up a postscript version of the text, including figures, via anonymous
%ftp to eku.ias.edu=192.16.204.30 in /pub/strauss/ors/paper1.ps.Z. The
%uncompressed file is 7 Mbytes, and even so, does not include Figure
%6, which can be taken as a separate file, fig6.ps.Z (4.5 Mbytes
%uncompressed). Send e-mail to strauss@guinness.ias.edu if you have
%problems.
\magnification=\magstep1
\hoffset=-0.6 true cm
\voffset=0.0 true cm
\baselineskip=20pt minus 1pt
\baselineskip=12pt

\headline={\hss\tenrm\folio\hss}
\footline={\hfil}
\vsize=8.9truein
\hsize=6.8truein
%\hyphenpenalty=10000
%\raggedright
%\parskip=\medskipamount
\tolerance=10000
\parindent=1truecm
\raggedbottom
\def\pp{\parshape 2 0truecm 5.8truein 2truecm 5.01truein}
\def\ltsima{$\; \buildrel < \over \sim \;$}
\def\simlt{\lower.5ex\hbox{\ltsima}}
\def\gtsima{$\; \buildrel > \over \sim \;$}
\def\simgt{\lower.5ex\hbox{\gtsima}}
\def\bline{\hbox to 1 in{\hrulefill}}
\def\etal{{\sl et al.\ }}

\def\np{\vfill\eject}

\def\kms{\ifmmode {\rm \ km \ s^{-1}}
\else
$\rm km \ s^{-1}$\fi}
\def\deg{^\circ}
\def\vp{\ifmmode {\bf r}
\else
{${\bf r}~$}\fi}
\def\self{\ifmmode {\phi ({\rm r})}
\else
{$\phi ({\rm r})~$}\fi}
\def\sfvp{\ifmmode {\phi ({\rm {\bf r}})}
\else
$\phi ({\rm {\bf r}})~$\fi}
\def\lumf{\ifmmode {\Phi ({\rm L,r})}
\else
{$\Phi ({\rm L,r})~$}\fi}
\def\lfvp{\ifmmode {\Phi ({\rm L,{\bf r}})}
\else
{$\Phi ({\rm L,{\bf r}})~$}\fi}
\def\n1{\ifmmode {{n_1}}
\else
${{n_1}}$\fi}

\def\iras{{\sl IRAS}}
\ \ \vskip 0.5 truein
\centerline{\bf THE OPTICAL REDSHIFT SURVEY:}
\medskip
\centerline{\bf SAMPLE SELECTION AND THE GALAXY
DISTRIBUTION\footnote{$^1$}{\rm Based in
part on data obtained at Lick Observatory, operated by
the University of California; the Multiple Mirror Telescope, a joint
facility of the Smithsonian Astrophysical Observatory and the
University of Arizona; Cerro Tololo Inter-American Observatory,
operated by the Association of Universities for Research in Astronomy,
Inc., under contract with the National Science Foundation; Palomar
Observatory, operated by the California Institute of Technology, the
Observatories of the Carnegie Institution and Cornell University;
Las Campanas Observatory, operated by the
Observatories of the Carnegie Institution.\vskip 1pt}}
\bigskip
\centerline{Bas\'\i lio X. Santiago\footnote{$^2$} {Institute of Astronomy,
Cambridge University, Madingley Road, Cambridge CB3 0HA, United
Kingdom\vskip 1pt},
Michael A. Strauss\footnote{$^3$} {Institute for Advanced Study,
School of Natural Sciences, Princeton, NJ 08540 U.S.A.\vskip 1pt}, Ofer
Lahav$^2$, }
\centerline{Marc Davis\footnote{$^4$} {Physics and Astronomy Departments,
University of California, Berkeley, CA 94720 U.S.A.\vskip 1pt}, Alan
Dressler\footnote{$^5$}{Observatories of the Carnegie Institution of
Washington, 813
Santa Barbara St., Pasadena, CA 91101 U.S.A.\vskip 1pt}, and John P.
Huchra\footnote{$^6$}
{Center for Astrophysics, 60 Garden St., Cambridge, MA 02138 U.S.A.}}
\bigskip
\bigskip
centerline{\it Submitted to The Astrophysical Journal}
\centerline{\it Received\ \hbox to 1in{\hrulefill};
accepted\ \hbox to 1in{\hrulefill}}
\bigskip
%\hrule\medskip
%\baselineskip=21pt
\np
\centerline{\bf ABSTRACT}
\medskip
This is the first in a series of papers describing the {\it Optical
Redshift Survey} (ORS), a redshift survey of optically selected
galaxies covering 98\% of the sky above $|b| = 20^\circ$
(8.09 ster). The
survey is drawn from the {\it Uppsala Galaxy Catalogue} (UGC), the {\it
European Southern Observatory Galaxy Catalogue} (ESO), and the {\it
Extension to the
Southern Galaxy Catalogue} (ESGC), and contains two sub-catalogs, one
complete to a $B$ magnitude of 14.5, the other complete to a $B$
major axis diameter of $1.9^\prime$. The entire sample consists of 8457
objects, of
which redshifts are now available for 8286; 171 objects remain without measured
redshifts. Roughly 1300 of the redshifts were measured for the
completion of the sample; the remainder were taken from the
literature. Most of these new redshifts are concentrated at low Galactic
latitudes
$20^\circ \leq |b| \leq 30^\circ$ and within the strip not covered by
either the UGC or ESO catalogues: $-17.5^\circ \leq \delta \leq -2.5^\circ$.
The ORS provides the most detailed and homogeneous sampling of the large-scale
galaxy distribution to date in these areas. The density field of bright optical
galaxies is well-defined to $8000 \kms$, and is
dominated by the Virgo, Hydra-Centaurus, Pisces-Perseus, Coma-A1367,
and Telescopium-Pavo-Indus Superclusters. The dense sampling provided by ORS
allows a detailed analysis of the galaxy density field, and will
be used to test its dependence
on morphology and other galaxy parameters.
\medskip
\noindent{\it Subject headings:} Cosmology: Large-Scale Structure of
Universe---Cosmology: Observations---Galaxies: Distances and Redshifts
\bigskip\goodbreak
%\np
\centerline {\bf 1. Introduction}
\medskip
Despite the immense increase in the number of galaxy redshifts measured in the
past
decade, there is much to learn from well-defined complete redshift
surveys of nearby galaxies.
The global mean density
of galaxies may still have significant uncertainties due to the
presence of large-scale structure comparable in size to the samples
themselves (de Lapparent \etal 1988; da Costa \etal 1994);
the measurement of various high-order galaxy clustering statistics are
limited by the finite volumes and sparse sampling of existing surveys
(Bouchet \etal 1993; Fisher \etal 1994b), and little is known
quantitatively about the relative distribution of galaxies of
different morphological types on very large scales.

The first complete redshift survey to cover both Galactic hemispheres
was that of Yahil, Sandage, \&
Tammann (1980), which mapped the Local
Supercluster using the galaxies of the Revised Shapley-Ames catalog
(Sandage \& Tammann 1981, hereafter RSA). Analyses of the optical
extragalactic sky were carried out using a compilation of the UGC,
ESO, and MCG catalogs in two dimensions (Lahav 1987; Lahav,
Rowan-Robinson, \& Lynden-Bell 1988); Lynden-Bell, Lahav, \&
Burstein (1989, hereafter LLB) and Hudson (1993a,b) used partial
redshift information with statistical corrections for incompleteness
to map the three-dimensional density field. When the database
of the {\it Infrared Astronomical
Satellite} (\iras) was produced, several groups used it as a target
list to carry out flux-limited, almost full-sky redshift surveys of
galaxies, which probed considerably beyond the Local Supercluster
(Strauss \etal 1990, 1992c; Fisher 1992; Rowan-Robinson \etal 1990;
Lawrence \etal 1994; Saunders \etal 1995). These surveys have been
used to trace the large-scale distribution of galaxies within
20,000 \kms\ of the Local Group (Saunders \etal 1991; Strauss \etal
1992a), and thus predict the peculiar velocity field via gravitational
instability theory (Yahil \etal 1991; Kaiser \etal 1991; Strauss \etal
1992b; Dekel \etal 1993; cf, Strauss 1993; Dekel 1994; Strauss \&
Willick 1995 for a review of scientific results from these surveys).

  The dynamical inferences from the \iras\ galaxy redshift surveys have
been called into question, however, because of our lack of knowledge
of the relative distribution of galaxies and mass. Moreover,
\iras\ galaxies are predominantly star-forming, dusty, late type
galaxies, and are under-represented in cores of
rich clusters relative to optically selected galaxies (Strauss \etal
1992a and references therein).  Indeed, on scales of 800 \kms, the rms
fluctuations in the \iras\ counts are smaller than those of optical
galaxies by a factor of 1.3--1.5 (e.g. Lahav, Nemiroff \&
Piran 1990; Scharf \etal 1992; Strauss \etal 1992a; Saunders,
Rowan-Robinson \& Lawrence 1992; Fisher \etal 1994a).

However, it is still unclear if these differences extend to larger
scales, and to what extent such differences affect various measures of
large-scale structure. For example, the dipole moment of the galaxy
density field can be compared with motion of the Local Group with
respect to the Cosmic Microwave Background in order to measure $\beta
\equiv \Omega^{0.6}/b$, where $\Omega$ is the cosmological density
parameter and $b$ is the bias factor. Lahav \etal (1988), Hudson
(1993a,b), Freudling \& da Costa (1994) and Hudson \& Dekel (1994)
have argued that $\beta$ for optical galaxies is smaller by a factor
of two than that derived from \iras\ galaxies. Strauss \etal (1992a)
argued that \iras\ and optically selected galaxies traced the same
large-scale structures, although Santiago \& Strauss (1992) found
significant difference between optically selected galaxies of
different Hubble types (cf., Lahav \& Saslaw 1992). Motivated by these
ambiguities, we undertook a new redshift survey of optically selected
galaxies covering most of the celestial sphere. We call it the {\it
Optical Redshift Survey}, or ORS. Our principal motivation is to
quantify any possible differences between the large-scale structure as
defined by \iras\ and optically-selected galaxies. In addition, we
wish to look for morphological segregation on large scales between
galaxies of different morphological types. Our hope is that this will
give new insights into the process of galaxy formation and possible
biasing of the galaxy distribution relative to that of mass on very
large scales, while testing many of the results based on the \iras\
density field.

The \iras\ redshift surveys have the advantage of an approximately
uniform selection over the entire sky (cf, the discussion in \iras\
1988; Strauss \etal 1990), while any optically-selected survey necessarily
is strongly affected by the zone of avoidance, and the presence of
extinction at low Galactic latitudes. In Santiago \etal (1994; hereafter
Paper~2), we will outline methods to correct for these problems.
However, the ORS is much more densely sampled than is the \iras\
survey, meaning that structures
can be delineated and quantified with less noise  on any given
smoothing scale. Moreover, the ORS includes the early-type galaxies
that are not represented in the \iras\ survey.

This is the first in a series of papers presenting results from the
ORS. We define the sample in \S 2, and in \S 3, we show the redshift
data using a variety of projections. Quantitative results from this
survey will be presented in a series of papers in preparation, as
outlined in \S 4. The data themselves will be published in the near
future.
\bigskip\goodbreak
\centerline {\bf 2. Sample Definition}
\medskip\nobreak
\centerline {\it 2.1 The catalogues used}
\medskip\nobreak
There exists no single galaxy catalogue selected from optical plate
material deeper than the RSA which covers both halves of the celestial sphere.
For
this reason, we have selected our redshift sample from three different
galaxy catalogues, as follows:
\medskip\nobreak
%\item {1-}
In the Southern Sky defined by
$\delta \leq -17.5 \deg$\footnote{$^7$}{All positions in this paper
are B1950.} we used the {\it ESO-Uppsala Survey of the ESO(B) Atlas}
(Lauberts 1982; ESO), as well as its photometric counterpart, the {\it
Surface Photometry Catalogue of the ESO-Uppsala Galaxies} (Lauberts \&
Valentijn 1989;
ESO-LV). The ESO catalogue is nominally complete in apparent visual
diameters, measured from the ESO(B) Survey plates down to $ \theta
\geq 1'$ (cf., Hudson \& Lynden-Bell 1991 [HLB], who show that the
completeness limit is closer to $1.35'$). The ESO-LV consists of the
results of photo-densitometer scans of these galaxies from the ESO(B)
and ESO(R) plates (although it is not complete; see below), and
tabulates magnitudes, diameters and surface brightnesses. ESO-LV also
contains data on a number of companion objects to the original ESO
galaxies, as well as on many
individual components of multiple systems of galaxies. The listed positions of
the
galaxies are accurate to $\sim 6^{\prime\prime}$ (Lauberts 1982).

%\item {2-}
In the Northern sky ($\delta \geq -2.5 \deg$),
galaxies were selected from the {\it Uppsala General Catalogue of Galaxies}
(Nilson 1973; UGC). This dataset is also
nominally complete for $\theta
\geq 1'$, as measured by eye from the {\it Palomar Sky Survey}
(POSS) 103a-O plates, although HLB show that the actual
completeness limit is closer to $\sim 1.65'$.
The UGC also includes all galaxies
with apparent $B$ band magnitude $m_B \leq 14.5$ as listed in the {\it
Catalog of Galaxies and Clusters of Galaxies} (Zwicky \etal 1961-1968;
CGCG). The CGCG is nominally complete to $m_B \leq 15.5$ and
covers the same area on the sky as the UGC, except for
regions close to the Galactic plane. The UGC galaxy positions are
of varying quality, and so we updated them with positions from the
NASA-Extragalactic Database (NED). In addition, we corrected
typographical errors in the ESO and UGC catalogues following Paturel
\etal (1991) and private communications with Dr.~Harold Corwin.

%\item {3-}
Galaxies in the strip between the ESO and UGC regions,
($-17.5 \deg \leq \delta \leq -2.5 \deg$) were sampled using the
recently completed {\it Extension to the Southern Galaxies Catalogue}
(Corwin \& Skiff
1994; ESGC). We hereafter refer to this region of sky as the ESGC
strip. The ESGC actually covers the region $-21 \deg \leq \delta \leq
3 \deg$ and is complete to major axis diameters of $1.9^\prime$ (see below) as
measured by eye on the POSS 103a-O plates. The positions in this catalogue are
accurate to a
few arc-sec in most cases; several positions were updated by BXS. An
alternate sample in this region of the sky is the {\it Morphological
Catalogue of Galaxies} (MCG) (Vorontsov-Velyaminov \etal 1962-1974),
but the photometry and
astrometry of this sample are known to be of poor quality. A previous
redshift survey of this region has been carried out by Pellegrini
\etal (1990) and Huchra \etal (1993).

 The completeness of the ESGC is demonstrated in Figure~1, which shows
the cumulative distribution of major axis diameters of the sample. In the
absence of clustering, the distribution is expected to be proportional
to $\theta^{-3}$ to the completeness limit. The dashed line shows this
expected distribution, which indeed is an excellent fit to for
diameters larger than $1.9^\prime$.
It fits the data poorly for diameters larger than $5^\prime$; this is
presumably due to large scale structure and shot noise. See HLB for
similar plots for the ESO and UGC samples.
\bigskip\goodbreak
\centerline {\it 2.2 Sample selection}
\medskip\nobreak
We wish to have a sample covering as much of the celestial sphere as
possible. Examination of the Galactic extinction maps of Burstein \&
Heiles (1982; BH) shows that there are few regions at Galactic
latitudes higher than $|b| = 20^\circ$ with excessive extinction, and
thus we decided to limit our survey at this latitude. In addition, we excluded
the regions
of the sky at higher latitude in which the $B$ band extinction according to BH
was greater than $A_B = 0.7$ mag (this excludes only 0.18 ster). Galaxies in
the
remainder of the celestial sphere were then selected from
the catalogues listed above, according to the union of
the following criteria:
\medskip
\item {1-} $\theta \geq 1.9'$, where $\theta$ is an apparent major
axis diameter.
For the ESO part of the sky, $\theta$ corresponds to the original
ESO B-band diameter rather than to any of the isophotal ESO-LV measurements,
as this is the quantity by which ESO was defined. We did not correct
diameters to face-on (cf., da Costa \etal 1988); major-axis diameters
are used throughout.
%For the UGC and ESGC regions, only one diameter measurement is available
%for each galaxy. {\bf I don't know what this sentence means!}
\item {2-} $m_B \leq 14.5$, where $m_B$ is an apparent magnitude
in the B band. For ESO, we used the ESO-LV total magnitudes,
$B_T$. Galaxies with ESO B diameters less than $1'$ were excluded,
even if they were brighter than the nominal flux limit, as the ESO
sample is incomplete below $1'$.
For UGC, CGCG magnitudes were used, as listed in the UGC
itself; for those objects in the UGC without CGCG magnitudes,
magnitudes were estimated by Nilson. There are no magnitudes (other than the
very poor ones from
the MCG) available in the ESGC strip. The CGCG magnitudes are known
to be systematically 0.3 magnitudes too faint in Volume 1 of the CGCG
($7^h < \alpha < 18^h, \ -2.5^\circ < \delta < 15^\circ$) (Kron \&
Shane 1976; Paturel 1977; Fasano 1985); we correct for this effect in
Paper~2.

For all parts of the sky, we select our sample according to {\it
observed} magnitudes or diameters as listed in the machine-readable
versions of the catalogues. In particular, in defining the sample
we do not correct the
magnitudes and diameters for Galactic or internal extinction, nor
do we apply corrections to bring the various photometric systems into
agreement (cf., LLB). It is known, for instance,
that the ESO diameter scale differs systematically from that of the UGC
(Lahav \etal 1988, HLB; see below).
Corrections for these effects will be applied in the analysis stage
by incorporating them in the sample selection function (Paper~2). However,
we can use the overlap between the samples to
compare the relative diameter scales, even if we will not use this
information here. Figure~2 compares the ESGC diameter scale with that of
ESO and UGC for those galaxies in common. Note the logarithmic axes. The
paucity of points below $D_{ESGC} = 112^{\prime\prime} = 1.9^\prime$ is just a
reflection of the completeness limit of this catalogue.
In each panel, the solid line
is $y = x$, while the dashed line is the best fit to $y = cx$. Thus we
find that:
  $$D_{ESO} = (0.918 \pm 0.014) D_{ESGC}\eqno(1)$$
and
  $$D_{UGC} = (0.735 \pm 0.010) D_{ESGC}\quad,\eqno(2)$$
implying that
  $$D_{ESO} = (1.249 \pm 0.026) D_{UGC}\quad.\eqno(3)$$
LLB determined this latter ratio by fitting
diameter functions to the ESO and UGC diameters; they obtained
$D_{ESO}/D_{UGC} = 1.17 \pm 0.07$, which is consistent with Equation 3
within the error bars. Note that
there is some evidence for a logarithmic slope steeper than unity in the
UGC-ESGC comparison; there may be non-linearities in the UGC diameter
scale. Other comparisons between different diameter systems, including
UGC and ESO, have been carried out by Fouqu\'e \& Paturel (1985),
Paturel \etal (1987) and de Vaucouleurs \etal (1991).

In Figure~3 we compare the original ESO diameters measured by eye to
those from ESO-LV measured at the $26^{th}$ isophotal level in the B
band. There is good agreement between the two diameter
scales, although at very small diameters (well below the ORS diameter
limit), the ESO-LV diameters are systematically larger.
%Because of the good correlation between
%isophotal and visual diameters, the ESO $D_{26} \ge 1.9'$ is still
%$\sim 95$\% complete in redshift. {\bf This sentence doesn't belong
%here, because we haven't even mentioned redshifts yet!}

%As we show in Paper~2 of this series, differences in the zero points,
%luminosity functions, or diameter functions from one catalog to
%another may be accounted for by computing a separate selection
%function for each individual dataset. Furthermore, the derived density
%field is quite insensitive to other systematic errors in the
%diameters or magnitudes as long as their amplitude does not vary across
%the sky. However, position-dependent errors (other
%than those caused by Galactic
%extinction, for which we can correct), are more problematic. In
%particular, the Zwicky magnitude zero-points are known to vary with
%declination (Fasano 1985, and references therein). Again, our
%selection is purely on the basis of quoted, observed magnitudes and
%diameters as given in the catalogs above; corrections for these
%effects will be carried out in the analysis stage, as described in
%Paper~2.

Another important issue in defining the ORS is the presence
of binaries or multiple systems of galaxies with a single entry in the
catalogs. These were excluded
from the sample, unless magnitudes or diameters were available
for individual components which satisfied the ORS selection criteria.
Information on individual galaxies in multiple systems is available
in the ESO-LV database; wherever the individual galaxies satisfied our
selection criteria, we included them in our sample. We searched the
notes section of the UGC for the 125 entries classified as multiple and whose
system diameters or magnitudes were large enough to make into ORS;
only seven of them had components whose individual magnitudes and
diameters caused them to enter our sample. For
the ESGC, separation of individual galaxies in multiples has not yet
been carried out, and we excluded the 59 entries classified as
unseparated multiples, and carefully checked the sample for duplicate
entries. We thank Harold Corwin for his help in sorting out
all doubtful cases.

Thus, we have two complete, although somewhat non-uniform, samples: a
diameter-limited sample covering the whole sky (other than those
regions for which $A_B > 0.7$) for $|b| > 20^\circ$, which we refer to as
ORS-d, and a magnitude-limited sample covering essentially the same
area minus the ESGC strip, which we refer to as ORS-m. We refer to the
corresponding subsamples covering only the ESO region of the sky as
ESO-d and ESO-m, and similarly for the UGC. As no magnitudes are
available for the ESGC region, we refer to its diameter-limited
subsample simply as ESGC.  There is of course very substantial overlap
between the ORS-m and ORS-d samples.

In Table 1 we provide information about the various ORS
subsamples: sample designations are given in column 1, selection criteria
are in column 2 and solid angles in column 3.
The total numbers of objects are given in column 4, while
the number and fraction with measured redshifts are
in columns 5 and 6, respectively. The final entry in Table 1
gives numbers for the union of ORS-m and ORS-d.

Unfortunately, ESO-LV magnitudes do not exist for the entire sky South
of $\delta < -17.5\deg$; there remain 64 plates that at least
partially cover the region $|b| > 20^\circ$ and that were not scanned.
Thus the ESO-m sample covers a smaller solid angle than does the ESO-d
(cf, Table 1).  In order to define the excluded zones unambiguously,
we exclude any ESO-m galaxy if its right ascension or declination
falls within $2.5^\circ$ of the center of one of these 64 plates, even
if it enters the catalog from an overlapping plate. Notice also that
there is a diameter selection inherent in ESO-m, as it does not
include galaxies brighter than $B_T = 14.5$ but smaller than the original
ESO diameter limit. The UGC-m sample does not suffer from this
problem: it is the union of a diameter-limited sample, and the subset
of the CGCG catalog with $m_B \leq 14.5$.

As the ORS sample as defined above consists of bright and/or large
galaxies, the majority had redshifts available in the literature, in
particular from the CfA (Huchra \etal 1983), CfA2 (Huchra \etal 1990),
SSRS (da Costa \etal 1991), SPS (Dressler 1991), MCG Equatorial (Huchra \etal
1993)
and \iras\ (Strauss \etal 1992c; Fisher 1992; Lawrence
\etal 1994) redshift surveys. In practice, we matched the catalog
against a private version of the redshift compilation of JPH
(ZCAT). This left $\sim 1300$ redshifts unmeasured, largely
concentrated in the regions of sky not previously covered by redshift
surveys: the ESGC strip, and the two strips defined by $20^\circ < |b|
< 30^\circ$. These redshifts were measured over four years at Palomar,
Lick, Mount Hopkins, Las Campanas, and Cerro Tololo Inter-American
Observatories. The typical error in these redshifts was 50 \kms.
The details of the observing runs and the reduction
procedures will be given in our data paper.

The completeness of the redshift survey is excellent: above 98\% for
all subsamples except for the ESGC strip (93.1\%). The
(diameter-limited) ESGC sample includes many very low surface
brightness galaxies for which it is very difficult to measure a
redshift. The ESO-d sample was defined in terms of visually estimated
diameters, but because of the good correlation with isophotal
diameters from ESO-LV (Figure~3), a sample defined by ESO $D_{26} \ge
1.9'$ is still $\sim 95$\% complete in redshift.

Figure~4 shows the redshift distribution of the different subsamples
normalized by their
respective solid angles. Each subsample shows
the effects of large-scale structure.
The UGC-m is noticeably deeper than is the
ESO-m sample. The peak at 1000 \kms\ in the UGC-m is due to the Ursa Major and
Virgo clusters. Six thousand \kms\ is roughly the redshift at which an
$L_*$ galaxy will be at the magnitude limit of the sample, causing the
peak there; the very sharp fall-off at larger distances is a
reflection of the exponential tail of the optical galaxy luminosity
function (Schechter 1976; Loveday \etal 1992, and references therein).
The ESGC is of comparable depth to the UGC-m, but is
more densely sampled by a factor of two (which may partly
be due to large-scale structure; the peak at 2000-3000 \kms\ is due to
the Great Attractor region). All three samples
cut off abruptly at 10,000 \kms. The UGC-d and ESO-d subsamples are also
shown; their cutoff at high redshift is not as strong as that for the
magnitude-limited subsamples, and the redshift distribution is much
flatter. In addition, the diameter-limited subsamples are quite a bit
more sparsely sampled than are the magnitude-limited subsamples.
For comparison, the \iras\ 1.2 Jy sample
(Fisher 1992) has a redshift distribution that peaks at a comparable
redshift to that of UGC-m, but has a high-redshift tail that extends
appreciably
further than do the optical surveys. This is a reflection of the
power-law tail of the \iras\ luminosity function at high luminosities,
in contrast with the exponential cutoff of the optical luminosity and
diameter functions. As discussed in the Introduction, the \iras\
sample is substantially sparser than are the optically selected
samples.
\bigskip\goodbreak
\centerline {\bf 3. The Distribution of Galaxies in the Survey}
\medskip\nobreak
\centerline{\it 3.1. Distribution on the Sky}
\medskip\nobreak
Before showing the sky distribution of the data themselves, Figure~5a
shows the geometry of the
survey in an Aitoff projection of the sky in Galactic coordinates. The
hashed region at low Galactic latitudes is the zone of avoidance, $|b|
< 20^\circ$, while the higher latitude black areas correspond to those
regions with $A_B > 0.7$. They occupy only 2\% of the sky outside of
the zone of avoidance, or 0.18 ster. The two largest such areas are in Orion
($l=150-190\deg,\ b=-30\deg$) and near the Galactic center ($l =
330-30\deg,\ b=20-30\deg$).  The grey areas at high Galactic
latitude are the plates not surveyed by ESO-LV; these represent
excluded zones for the ESO-m sample, but not the ESO-d. They cover an
additional 0.33 ster, or 4\% of the high-latitude sky. The two
S-shaped lines are drawn at $\delta=-2.5^\circ$ and
$\delta=-17.5^\circ$, respectively, and mark the boundaries between
the ESO, ESGC, and UGC samples. Figure~5b shows the distribution of
those galaxies for which we were unable to measure redshifts. The
highest concentration of these is in the ESGC strip, again indicated
with two S-shaped lines. We attempted to measure redshifts for most of
these objects, but the majority of them were of very low surface
brightness. In the ESO and
UGC regions, most of the galaxies without redshifts are in the diameter-limited
sample, and are also of very low surface
brightness. There were also some galaxies with bright stars superposed for
which
we were unable to obtain a spectrum.

In order to present the data themselves, we use the union of the
UGC-m, ESO-m, and ESGC data. The description here will be qualitative;
more systematic and quantitative analysis of the data will be
presented in the upcoming papers of this series.

Figure~6a shows the distribution of the entire sample on the celestial sphere
using an equal area polar hemispheric projection in Galactic
coordinates (Lahav 1987, Scharf \& Lahav 1993). The circle on the left
(right) side corresponds to the
North (South) Galactic hemisphere. The poles are at the center of
these circles with Galactic latitude decreasing radially
outwards; the dashed circles are drawn at Galactic latitudes
$20^\circ,\ 40^\circ,\ 60^\circ$, and $80^\circ$ in the North, and
$-20^\circ,\ -40^\circ,\ -60^\circ$ and $-80^\circ$ in the South.
Galactic longitude runs azimuthally as indicated. The
circles are rotated so as to make the Supergalactic Plane (SGP)
stretch along the horizontal direction.  The shaded regions correspond
to the the excluded zones (compare with Figure~5a).
The ESGC strip is marked with two solid lines in both hemispheres.
The outer ring devoid of galaxies is the zone of avoidance ($|b| <
20^\circ$).

Interpretation of structures in this map is difficult: each
subsample making up the survey has a different selection function
(Figure~4), large-scale Galactic extinction creates
artificial low-density regions, and the projection of all the data tends to
wash out
features. However, a few structures catch the eye immediately:
the densest part of the Local Supercluster and the Hydra-Centaurus
Supercluster is prominent in the Northern Hemisphere at $l =
300-315^\circ$ from $b = 20^\circ$ almost to the North Pole. It appears to
be contiguous with the Telescopium-Pavo-Indus Supercluster (TPI) in the
Southern Hemisphere. The Pisces-Perseus chain is also visible at $l =
105-135^\circ$ and $b = -30^\circ$.

Figure~6b shows the same data, now in an Aitoff projection. This
complementary projection of the data makes clearer the relation of
structures to one another near the Galactic Plane, but of course distorts
features
near the poles. Again, the plane of the Local Supercluster, the TPI
Supercluster, and Pisces-Perseus Supercluster are all apparent. The
lower density of galaxies at small Galactic longitudes is largely due
to Galactic extinction.

In order to distinguish structures more clearly, we make use of our
redshift data, plotting the galaxies in constant redshift slices. For
each slice, we show two figures, one in hemispherical projection, as
in Figure~6a, and one in Aitoff projection as in Figure~6b. In each
case, we also show the galaxy density field on a constant redshift shell,
smoothed with a Gaussian whose width increases proportional to the
mean interparticle separation. This analysis corrects for extinction,
the excluded zones, the selection function of each catalog, and other
related effects; details are in Paper~2.

The upper panels of Figures~7a and 7b show the galaxy distribution of
galaxies with redshifts (corrected
for the motion of the Sun relative to the barycenter of the Local
Group following Yahil, Tammann, \& Sandage 1977) between 0 and 3000
\kms. The lower panels show the smoothed galaxy density field on the
spherical shell with radius 1500 \kms. The heavy contour is at the
mean density, the dotted contours are at $1/3$ and $2/3$ the mean
density, and the solid contours represent positive
density contrasts with logarithmic spacing; three contours are a
factor of two overdense. Note that the coordinate grid used in the upper
panel of Figure~7a is not included in the lower panel, for clarity.
The labels refer to various structures visible in these plots.
%For the southern half of the sky we refer the reader to Pellegrini
%\etal (1990) for an excellent description of the regions
%below $b = -30^\circ$.
The densest part of the Local Supercluster is the strong overdensity at $l
=300-315^\circ$, $b = 30-70^\circ$; the Virgo cluster (V) is at its
Northern tip. The Ursa Major cluster ($l = 140^\circ,\ b = 60^\circ$,
UM) is also visible, also in the Supergalactic plane; indeed, galaxies
are concentrated to the Supergalactic plane all across the Northern
hemisphere. Several other noticeable clumps belong to the Local Supercluster,
among them
the Leo ($l = 225^\circ,\ b=60^\circ$), Virgo-Libra ($l=345^\circ,\
b=35^\circ$) and Canes Venatici-Camelopardalis ($l=95^\circ$,
$50^\circ < b < 70^\circ$) clouds (cf., Tully \& Fisher 1987). All
these clouds are too small to appear in the contour plots at this
smoothing.
The Dorado-Fornax-Eridanus complex (DFE; the Southern Supercluster of
Mitra 1989) ranges from $l
= 190^\circ$, $b = -30^\circ$ to  $l = 270^\circ$, $b
= -40^\circ$, and the foreground of the TPI Supercluster appears at
$330^\circ \leq l \leq 15^\circ$, $-50^\circ \leq b \leq -20^\circ$.
The Local Void (LV) of Tully \& Fisher (1987) is reflected in the paucity
of galaxies at low Galactic longitudes. There is a void at $l =
285^\circ$,$b = -60^\circ$ named  V1 by
da Costa \etal (1988) visible in the galaxy distribution; its center
is at 2500 \kms, and thus it is not apparent in the contour plots.
Detailed maps of these foreground structures,
in both hemispheres, can be found in Tully \& Fisher (1987). See also
Pellegrini \etal (1990) for a detailed discussion of the cosmography
in the Southern Hemisphere.

%In the north, the SGP stands out clearly in Figure~8b as a dense
%horizontal structure. The Virgo cluster and its southern extension
%are visible as a very high density region at $285^\circ \leq l \leq
%%320^\circ$,
%$30^\circ \leq b \leq 75^\circ$.
%The Ursa Major cloud ($l \sim 140^\circ$, $b \sim 60^\circ$) is also
%clearly visible.
%The presence of all these foreground structures makes the
%northern panel much denser than its southern counterpart. {\bf Not
%really true!.}

Figures 8a and 8b show the galaxy distribution in the slice from 3000
to 6000 \kms, and the smoothed density field at 4500 \kms. This is the
redshift interval which the ORS samples most completely (Figure~4). In
particular, this slice
intersects the Great Attractor, which includes the TPI supercluster
stretching from $l = 0^\circ$, $b = -20^\circ$ to $l = 30^\circ$, $b =
-60^\circ$, and the Hydra-Centaurus Supercluster (Hydra (H): $l =
255-275^\circ$, $b = 30^\circ$; Centaurus (Cen): $l =
300-320^\circ$, $b = 25-45^\circ$).  The Pisces-Perseus
Supercluster (PP), studied in detail by Haynes \& Giovanelli (1988), is
on the opposite side of the sky at $l = 110-150^\circ$, $-35^\circ
\leq b \leq -20^\circ$, and the N1600 group (represented by an N in
the figures) at $l = 200^\circ$, $b = -25^\circ$. The latter is
within the ESGC strip. There is some evidence for a connection between
the PP and TPI superclusters, especially visible in Figure 8a. The Cetus wall
(Ce) stretches from $l =
120^\circ$ through $l = 180^\circ$, to the West of the Southern
Galactic pole. The void at $l = 315^\circ,\ b=-45^\circ$ was called V3
by da Costa \etal 1988.
The Northern sky is dominated by filamentary patterns and voids (Davis
\etal 1982).  The Camelopardalis-A569 ($l = 145^\circ$, $b =
30^\circ$, CAM) region is an overdensity which may be contiguous with the
Pisces-Perseus Supercluster (Hudson 1992), and the Cancer cluster ($l
= 195^\circ, b = 25^\circ$, CAN) is at the edge of the zone of
avoidance. The extensive void covering most of the Northern Galactic
hemisphere between $l=45^\circ$ and $l=195^\circ$ lies between the
Virgo cluster and the great Wall, and was apparent in the CfA survey
of Davis \etal (1982).

%{\bf Michael, that is where I got
%stuck with the northern sky. Also clearly
%visible are: the long filament crossing the pole and bending to
%the right towards the Centaurus region, which is in the
%foreground of the Coma-A1367 supercluster and the Great Wall; the
%filament spanning from $l = 350^\circ$ to $l = 75^\circ$, $b =
%60^\circ$; the clump at $\l = 90^\circ$, $b = 35^\circ$. What are
%these things? The second filament I mention seems to be associated to
%a cluster, MKW12 (see Hudson 1992).
%The void to the right of the clump and south of this
%second filament is V4 of Saunders \etal 1991. That is all I know by now.}

Figures 9a and 9b show the galaxy distribution from 6000 to 9000 \kms,
and the density field at 7500 \kms. The ORS is fast becoming sparse at
these redshifts, and individual structures are difficult to discern.
However, the Coma cluster and Abell 1367 (labeled Co) make up the overdensity
centered on the North Galactic pole, and the structure of which this
is a part, extending to $l = 245^\circ$, $b = 55^\circ$, and $l =
355^\circ$, $b = 65^\circ$, is the southern extension of the Great
Wall (GW) of de Lapparent, Geller,
\& Huchra (1986; 1988) and Geller \& Huchra (1989), where it dips down
to low redshifts. Saunders \etal (1991) defined the Supercluster S5
at $l = 95^\circ$, $b = -30^\circ$.

%A smoothed version of Figure~7 is shown in Figure~8. The contours
%shown correspond to isodensity lines, in steps of 0.2 dex. The panels
%in Figure~8 correspond to the same redshift shells shown in the last 3
%panels of Figure~7.  The densities were computed in a grid of points
%located at the center of each redshift shell considered.  A gaussian
%smoothing was applied to the discrete galaxy distribution.  Each
%galaxy was weighted by the inverse of the selection function at its
%position. The selection function was computed as explained in Paper~2
%and takes into account the presence of variable amounts of Galactic
%extinction.  The heavy contour indicates the mean density level.
%Solid (dashed) lines correspond to densities above (below) the
%mean. Labels for the main clusters and superclusters are shown in
%order to facilitate their identification.  Figure~8 provides a picture
%more exempt of selection effects than that of Figure~7. A more
%quantitative analysis of the galaxy density field is going to be
%presented in an upcoming paper of this series.
\bigskip\goodbreak
\centerline{\it 3.2. Distribution in Radius}
\medskip\nobreak
As mentioned above, most of the new redshifts for this project were
measured in regions of the sky not previously surveyed: the ESGC
strip, and the two strips defined by $20\deg < |b| < 30\deg$. Thus we
discuss the galaxy distribution in these three regions in some detail.

Figures 10 and 11 show the strips with $-30 \deg \leq b \leq -20 \deg$
and $20 \deg \leq b \leq 30 \deg$, respectively. These strips are
parallel to the zone of avoidance and so span $360 \deg$ in the sky.
Galactic longitude is the angular coordinate in these plots; Galactic
latitude has been suppressed.  Our Local Group is in the center of
each plot and redshift distance runs radially towards the edge. The
velocities are again in the Local Group frame; no further corrections
for peculiar velocities were applied. The dashed circles have radii of
2000, 4000, and 6000 \kms; the outer circle has a radius of 8000
\kms. Regions heavily affected by Galactic extinction are indicated
by the pie-shaped wedges (compare with Figure~5a).

Figure~10 is dominated by 2 superclusters:
the Pisces-Perseus Supercluster spanning the range  $125 \deg \leq l \leq 155
\deg$ and with redshifts ranging from 4000 to 6000 \kms, and the
Telescopium-Pavo-Indus Supercluster, concentrated at $l=330\deg$ and
4000 \kms, and with an extension between 4000 and 6000 \kms \  to $l =
0\deg$.
The galaxy concentration at $l = 200 \deg$, $v = 4500 \kms$ is the
N1600 region (Saunders \etal 1991), while the extended
structure at $l = 225 \deg$,
$v = 3500 \kms$ is a supercluster named S1 by Saunders \etal
(1991). Another of these superclusters, named S5,
is seen at $l = 95 \deg$, $v = 7000 \kms$ and seems to be connected
to the PP Supercluster.

The galaxy distribution in Figure~11
is also quite non-uniform, with galaxies concentrated in
large lumps surrounded by voids. The figure is dominated by the
Virgo-Hydra-Centaurus region, which occupies
the interval $260 \deg \leq l \leq 310 \deg$ and $2000 \kms \leq v
\leq 4500 \kms$. The clump at $l=270\deg$, 3000 \kms\ is Abell 1060,
and that at $l=330\deg$, 4000 \kms\ consists of Abell 3526, 3560, and
3565 in the Hydra-Centaurus region. Also visible are the
Abell 569 supercluster
($l = 170 \deg$, $v = 6000 \kms$), the Cam supercluster
($l = 135 \deg$, $v = 4500 \kms$) and the Cancer cluster
($l = 195 \deg$, $v = 4500 \kms$) (see also Hudson 1992).
The concentration of galaxies
to the Supergalactic plane (crossing the panel from $l = 135^\circ$ to
$l = 315^\circ$) is quite dramatic: notice the large void at $l
= 260 - 300\deg$ and redshifts beyond 2000 \kms, as well as the paucity
of galaxies at $l=70-120\deg$ at all redshifts in this slice.
%Given that few redshift surveys
%in the optical have probed this close to the Galactic plane,
%some of the smaller galaxy associations seen in
%Figures 9 and 10 are perhaps being noticed for the first time. {\bf
%This sentence adds nothing.}

In Figure~12 we show the ESGC strip. The two nearly opposite
regions completely devoid of galaxies are the intercepts with the
plane of the Galaxy.
As we saw in Figure~4, the ESGC strip is more densely sampled than is
the rest of the sky, making dense regions stand out more clearly than
in Figures 10 and 11.
The radially extended feature at $\alpha = 13^h$ is a piece of the
Virgo-Hydra-Centaurus complex, particularly by the Centaurus group ($v
= 2500 \kms$) and the Virgo Southern extension ($v = 1000
\kms$). Opposite it ($ 2^h \leq \alpha \leq 3^h$, $v = 2000 \kms$) is
the continuation of the Fornax-Eridanus complex towards the
Equator. Nearly in the same direction but in the background,
there is a very extended linear feature ranging from $\alpha = 0^h$
to $\alpha = 4^h$. This is the Northern extension of the Cetus Wall,
as mapped by the Southern Sky Redshift Survey (da Costa \etal 1988)
and the MCG Equatorial Survey (Huchra \etal 1993); cf., da Costa \etal
(1994). The latter authors hypothesize that this structure is
contiguous with the Pisces-Perseus Supercluster. In fact, this connection
can also be seen in Figure~8. To the East, there is
a prominent structure at $\alpha = 5^h$, but in fact this is not to be
identified with any single cluster; although it is close to N1600,
it is spread out over some 100 square degrees.
\bigskip\goodbreak
\centerline {\bf 4. Discussion }
\medskip\nobreak
In this paper we have presented a sample of optical galaxies covering most
of the celestial sphere and with almost complete redshift information.
The ORS provides a dense sampling of the structures that dominate
the density field in the Local Universe out to 8000 \kms. Among the
new areas surveyed by ORS are the regions close to the plane of the Galaxy
and the ESGC strip. The galaxy distribution in these regions is shown
in greater detail and completeness than in earlier surveys. In
particular, the Virgo Supercluster, the Doradus-Fornax-Eridanus
complex, Hydra-Centaurus Supercluster,
the Telescopium-Pavo-Indus Supercluster, and the low-redshift
component of the Great Wall are all
apparent in the data.
%The data show a clear continuation of the main superclusters in the
%local universe such as the Virgo-Hydra-Centaurus region,
%Perseus-Pisces and Pavo-Indus-Telescopium. Some previously observed
%features and patterns of the large-scale galaxy distribution are
%confirmed here with more detail.  In particular, we observe a
%high-density, radially extended region close to the ESGC strip at low
%galactic latitude, possibly associated to the N1600 cluster.

Given its high density sampling and large sky coverage, the ORS is well
suited for a quantitative analysis of the density field as traced by
bright optical galaxies and of its dependence on various
parameters. Particularly important will be the comparison of the ORS
density field with that described by \iras\ galaxies, to address the
question whether determinations of the cosmological density parameter
from these samples are consistent. Utilization of a
composite \iras-optical sample may be useful for deriving the peculiar
velocity field and constraining the bias and density parameters.

A density field analysis of optical galaxies
has recently been carried out by Hudson (1993a) who used the
entire ESO and UGC catalogues but with incomplete redshift
information. In addition, unlike the present sample, Hudson had no
information about the galaxy distribution in the ESGC strip.
The density field described here is qualitatively similar to that
presented by Hudson. He finds that optically selected galaxies show
stronger clustering than do \iras\ galaxies by a factor of $\sim 1.5$
on 800 \kms\ scales; some indication for morphological segregation is
also observed.

Another optically selected redshift survey has been carried out by
Freudling \& da Costa (1994). This sample contains about 5800 objects
limited to $m_B \leq 14.5$ in the regions $b \geq 40 \deg$ or $b \leq
-30 \deg$. They are making models for the peculiar velocity field as
inferred from this optical sample in combination with \iras\ (cf.,
Freudling \etal 1994).

There are two developments on the horizon that should allow the
selection criteria of the present sample to be tightened considerably.
B. Madore \& G. Helou are carrying out a photometric CCD survey in $R$
band of 10,000 galaxies based on the sample selection criteria here.
These data will give accurate diameters and magnitudes for all
galaxies in the present survey, allowing a much tighter selection. In
addition, D. Lynden-Bell and collaborators have carried out an
APM galaxy survey
in the ESGC strip that will be complete to $\theta_B \geq 0.4^\prime$ or
$m_B \leq 17$; all galaxies have been scanned from the
APM plates, yielding a database as rich as that of ESO-LV, but extending
substantially deeper.

ORS is affected by variable amounts of galactic absorption
across the sky. Systematic selection effects as a function
of direction in the sky may also arise due to
inconsistencies among the different
magnitude and diameter systems used in defining the sample (cf.,
Figure~2). We present methods for correcting for these effects, and
creating a well-defined density field, in Paper~2 of this series.
Once this is done, the scientific questions which
we hope to address with the present
sample include:
\medskip
\item{1-} Are there significant variations in the large-scale galaxy
distribution as a function of morphology, spectral properties or
surface brightness? In particular, are there significant differences
in the density fields of \iras\ and optical galaxies?
\item{2-} What is the amount of relative biasing in the galaxy distribution
and how does it depend on scale? Is relative biasing non-linear?
\item{3-} What are the high-order clustering properties of the galaxy
distribution? More specifically, do they confirm the scale-invariant
hypothesis and, if so, on what scales?
\item{4-} What is the dimensionality, density contrast
and extent of the Local Supercluster?
\medskip\nobreak
We intend to tackle all the above questions in upcoming papers.
\bigskip
{\bf Acknowledgments.} Harold Corwin deserves special
thanks for his enormous efforts in
producing the ESGC, making it available to us well in advance of
publication, and for much help and useful advice throughout
this project. We would also like to thank Barry Madore, Donald
Lynden-Bell and Dave Burstein for their advice in many aspects
of this project. Nicolaci
da Costa was particularly generous with redshifts in advance of
publication. Amber Miller assisted with the data reduction. Robert
Lupton wrote the software package used to generate the figures, and
cheerfully made improvements in the code when we requested them.
The telescope operators of Lick, Mount Hopkins, Las
Campanas, Palomar, and Cerro Tololo Observatories all did their usual
superb jobs. MAS is supported at the IAS under NSF grant
\#{}PHY92-45317, and grants from the W.M. Keck Foundation and the Ambrose
Monell Foundation. MD acknowledges support of NSF grant AST-9221540.
This research has made use of the NASA/IPAC Extragalactic Database (NED)
which is operated by the Jet Propulsion Laboratory, Caltech, under contract
with the National Aeronautics and Space Administration.
BXS acknowledges a doctoral fellowship from the
{\it Conselho Nacional de Desenvolvimento Cient\'\i fico e Tecnol\'ogico}
(CNPq), and the generous hospitality of the
Caltech Astronomy department. BXS and OL thank the Institute for
Advanced Study for its invitations to visit.
\vfill\eject
\centerline {{\bf Table 1}. ORS subsamples }
\bigskip
\hrule
\tabskip=1em plus2em minus.5em
\halign to\hsize
{#\hfil&\hfil#&\hfil#&\hfil#&\hfil#&\hfil#\cr
\noalign{\smallskip}
Catalogue &Cut-off &Solid Angle (ster) &\# of objects & \# with $z$ &
Completeness \cr
\noalign{\smallskip\hrule\smallskip}
 ESO-d & $\theta \geq 1.9'$ &2.70 & 1845 & 1821 & 0.987 \cr
 ESO-m & $B_T \leq 14.5$ &2.37 & 2437 & 2425 & 0.995 \cr
 ESGC & $\theta \geq 1.9'$ &1.14 & 1454 & 1353 & 0.931 \cr
 UGC-d & $\theta \geq 1.9'$ &4.25 & 2035 & 2002 & 0.984 \cr
 UGC-m & $m_B \leq 14.5$ &4.25 & 3279 & 3272 & 0.998 \cr
\noalign{\medskip}
 ORS-m & $m_B \leq 14.5$ & 6.62 & 5716 & 5697 & 0.997 \cr
 ORS-d & $\theta \geq 1.9'$ & 8.09 & 5334 & 5176 & 0.970 \cr
 ORS-m${}\cup{}$ORS-d & & 8.09 & 8457 & 8286 & 0.980 \cr}
\smallskip\hrule\smallskip

\vfill\eject
\centerline {\bf REFERENCES}
\def\pp{\parshape 2 0truecm 13.4truecm 2.0truecm 11.4truecm}
\def\apjref #1;#2;#3;#4 {\par\pp#1, #2, #3, #4 \par}
\def\apj{\apjref}
\medskip
\hyphenation{MNRAS}
%\sevenrm

\par\pp Bouchet, F.R., Davis, M., Strauss, M.A., Fisher, K.B., Yahil, A.,
\& Huchra, J.P. 1993, ApJ, 417, 36

\par\pp Burstein, D., \& Heiles, C. 1982, AJ, 87, 1165

\par\pp Corwin, H.G, \& Skiff, B.A. 1994, Extension to the Southern Galaxies
Catalogue, in
preparation

\apjref da Costa, L.N., Pellegrini, P., Davis, M., Meiksin, A.,
Sargent, W., \& Tonry, J. 1991;ApJS;75;935

\apjref da Costa, L. N. \etal
%, Pellegrini, P., Sargent, W., Tonry, J., Davis, M.,
% Meiksin, A., Latham D., Menzies, J., \& Coulson, I.
1988;ApJ;327;544

\par\pp da Costa, L.N. \etal 1994, ApJ, 424, L1

%\apj Davis, M., Geller, M.J, \& Huchra, J. 1978;ApJ;221;1

\par\pp Davis, M., Huchra, J., Latham, D.W., \& Tonry, J. 1982, ApJ, 253, 423

\par\pp Dekel, A. 1994, ARAA, in press

\par\pp Dekel, A., Bertschinger, E., Yahil, A., Strauss, M.A., Davis, M.,
\& Huchra, J.P. 1993, ApJ, 412, 1

\apj de Lapparent, V., Geller, M.J., \& Huchra, J.P.
1986;ApJ;302;L1

\apjref de Lapparent, V., Geller, M.J., \& Huchra, J.P. 1988;ApJ;332;44

\pp de Vaucouleurs, G., de Vaucouleurs, A., Corwin, H., Buta, R., Paturel,
G, \& Fouqu\'e, P., 1991. {\sl Third Reference Catalogue of Bright
Galaxies}, Springer-Verlag, New York

\par\pp Dressler, A. 1991, {ApJS}, 75, 241

\pp Fasano, G., 1985, A\&AS, 60, 285

\par\pp Fisher, K.B. 1992, PhD Thesis, University of California, Berkeley

\par\pp Fisher, K.B., Davis, M., Strauss, M.A.,  Yahil, A., \& Huchra,
J.P. 1994a, MNRAS, 266, 50

\par\pp Fisher, K.B., Davis, M., Strauss, M.A.,  Yahil, A., \& Huchra,
J.P. 1994b, MNRAS, 267, 927

\apj Fouqu\'e, P., \& Paturel, G. 1985;A\&A;150;192

\par\pp Freudling, W., \& da Costa, L.N., 1993, in {\it Cosmic Velocity
Fields}, eds: F.R. Bouchet \& M. Lachi\`eze-Rey (Gif sur Yvette: Editions
Fronti\`eres), 187

\par\pp Freudling, W., da Costa, L.N., \& Pellegrini, P.S. 1994,
MNRAS, in preparation

\apj Geller, M.J., \& Huchra, J.P. 1989;Science;246;897

\par\pp Haynes, M.P., \& Giovanelli, R. 1988,
in {Large Scale Motions in the
Universe: A Vatican Study Week}, edited by V.C. Rubin \& G.V. Coyne, S.J.
(Princeton: Princeton University Press), p.\ 31

\apj Huchra, J., Davis, M., Latham, D., \& Tonry, J. 1983;ApJS;52;89

\pp Huchra, J., Geller, M.J., de Lapparent, V. \& Corwin, H., 1990,
ApJS, 72, 433

\par\pp Huchra, J., Latham, D.W., da Costa, L.N., Pellegrini, P.S., \&
Willmer, C.N.A. 1993, AJ, 105, 1637

\par\pp Hudson, M.J., 1992, PhD thesis, Univ. Cambridge

\par\pp Hudson, M.J., 1993a, MNRAS, 265, 43

\par\pp Hudson, M.J., 1993b, MNRAS, 266, 475

\par\pp Hudson, M.J., \& Dekel, A. 1994, in preparation

\par\pp Hudson, M.J., \& Lynden-Bell, D. 1991, MNRAS, 252, 219 (HLB)

\par\pp {{{\sl IRAS} Catalogs \& Atlases, Explanatory Supplement} 1988,
edited by C.\ A.\ Beichman, G.\ Neugebauer, H.\ J.\ Habing, P.\ E.\ Clegg, \&
T.\ J.\ Chester (Washington D.C.:\ U.S. Government Printing Office)}

%\par\pp {{\sl IRAS} {Point Source Catalog} Version 2.\ 1988, Joint
%{\sl IRAS} Science
%Working Group (Washington D.C.:\ U.S. Government Printing Office) (PSC)}

\apjref Kaiser, N., Efstathiou, G., Ellis, R., Frenk, C., Lawrence,
A., Rowan-Robin\-son, M., \& Saunders, W. 1991;MNRAS;252;1

\par\pp Kron, G.E., \& Shane, C.D. 1976, A\&SS, 39, 401

\apjref Lahav, O. 1987;MNRAS;225;213

\apj Lahav, O., Nemiroff, R.J., \& Piran, T. 1990;ApJ;350;119

\apjref Lahav, O., Rowan-Robinson, M., \& Lynden-Bell, D. 1988;MNRAS;234;677

\pp Lahav, O., \& Saslaw, W., 1992, ApJ, 396, 430

\par\pp Lauberts, A. 1982, {The ESO-Uppsala Survey of the ESO(B) Atlas},
(European Southern Observatory)

\par\pp Lauberts, A., \& Valentijn, E. 1989, {\it The Surface Photometry
Catalogue of the ESO-Uppsala Galaxies}, (European Southern Observatory)
(ESO-LV)

\par\pp Lawrence, A. \etal 1994, MNRAS, in preparation

\apj Loveday, J., Peterson, B.A., Efstathiou, G., \& Maddox, S.J.
1992;ApJ;390;338

\apj Lynden-Bell, D., Lahav, O., \& Burstein, D. 1989;MNRAS;241;325 (LLB)

%\apj Maddox, S.J., Efstathiou, G., Sutherland, W.J., \& Loveday, J.
%1990;MNRAS;242;43P
%
\apj Mitra, S. 1989;AJ;98;1175

\par\pp Nilson, P. 1973, {Uppsala General Catalogue of Galaxies}, {
Uppsala Astron.\ Obs.\ Ann.}, {6}

\par\pp Paturel, G. 1977, A\&A, 56, 259

\apj Paturel, G., Fouqu\'e, P., Lauberts, A., Valentijn, E.A., Corwin, H.G.,
\& de Vaucouleurs, G. 1987;A\&A;184;86

\par\pp Paturel, G., Petit, C., Kogoshvili, N., Dubois, P.,
Bottinelli, L., Fouqu\'e, P., Garnier, R., \& Gouguenheim, L. 1991,
A\&AS, 91, 371

\par\pp Pellegrini, P.S., da Costa, L.N., Huchra, J.P., Latham, D.W.,
and Willmer, C. 1990, AJ, 99, 751

\pp {Rowan-Robinson, M., Lawrence, A., Saunders, W., Crawford, J., Ellis, R.,
Frenk, C.~S., Parry, I., Xiaoyang, X., Allington-Smith, J., Efstathiou, G., \&
Kaiser, N. 1990, MNRAS, 247, 1}

\par\pp Sandage, A., \& Tammann, G.A. 1981, {Revised Shapley-Ames
Catalog}, (Washington: Carnegie Institute of Washington) (RSA)

\par\pp Santiago, B.X., \& Strauss, M.A. 1992, ApJ, 387, 9

\par\pp Santiago, B.X. \etal 1994, in preparation (Paper~2)
%, Strauss, M. A., Lahav, O., Davis, M., Huchra, J., \& Dressler, A., 1994, in
%%preparation.

\par\pp {Saunders, W., Frenk, C.~S., Rowan-Robinson, M., Efstathiou, G.,
Lawrence, A.,  Kaiser, N., Ellis, R.~S., Crawford, J., Xia, X.-Y., \&
Parry, I. 1991, Nature, 349, 32}

\pp Saunders, W., Rowan-Robinson, M., Lawrence, A. 1992, MNRAS, 258,
134
%\pp {Saunders, W., Rowan-Robinson, M., Lawrence, A., Efstathiou, G., Kaiser,
%N., Ellis, R.~S., \& Frenk, C.~S.~1990, MNRAS, 242, 318}
%
%
\par\pp Saunders, W. \etal 1995, in preparation.

\par\pp Scharf, C., Hoffman,Y., Lahav, O., \& Lynden-Bell, D. 1992,
MNRAS, 256, 229

\par\pp Scharf, C., \& Lahav, O., 1993, MNRAS, 264, 439

\apj Schechter, P.L. 1976;ApJ;203;297

%\par\pp {Soifer, B.\ T., Sanders, D.\ B., Madore, B.\ F., Neugebauer, G.,
%Danielson, G.\ E., Elias, J.\ H., Lonsdale, C.\ J., \& Rice, W.\ L.\ 1987,
%{ApJ}, {320}, 238}
%

\par\pp Strauss, M.A. 1993, in {\it Sky
Surveys: Protostars to Protogalaxies}, edited by B.T. Soifer,
ASP Conference Series \# 43, 153

\par\pp Strauss, M.A., Davis, M., Yahil, A., \& Huchra, J.P. 1990, {ApJ},
{361}, 49

\par\pp Strauss, M.A., Davis, M., Yahil, A., \& Huchra, J.P. 1992a, {ApJ},
385, 421

\par\pp Strauss, M.A., Huchra, J.P., Davis, M., Yahil, A., Fisher,
K.B., \& Tonry, J.P. 1992c, ApJS, 83, 29

\par\pp Strauss, M.A., \& Willick, J. 1994, Physics Reports, in preparation

\par\pp Strauss, M.A., Yahil, A., Davis, M., Huchra, J.P., \& Fisher,
K.B. 1992b, {ApJ}, 397, 395

\par\pp Tully, R.B., \& Fisher, J.R. 1987, {Nearby Galaxies Atlas}
(Cambridge: Cambridge University Press)

\par\pp Vorontsov-Velyaminov, B.A., Archipova, V.P., \& Krasnogorskaja, A.A.,
1962--1974, {\it Morphological Catalogue of Galaxies},
in five volumes, (Moscow: Moscow State University) (MCG)

\apj Yahil, A., Tammann, G.A., \& Sandage, A. 1977;ApJ;217;903

\apjref Yahil, A., Sandage, A., \& Tammann, G.A. 1980;ApJ;242;{448}

\par\pp Yahil, A., Strauss, M.A., Davis, M., \& Huchra, J.P. 1991,
{ApJ}, 372, 380

\par\pp Zwicky, F. \etal 1961--1968, {\it Catalog of Galaxies and Clusters of
Galaxies}, in six volumes (Pasadena: California Institute of
Technology) (CGCG)

\vfill\eject

\def\fig #1, #2, #3 {\par\pp #3}
\centerline {\bf FIGURE CAPTIONS}
\medskip

\fig esgcdiam.ps, 6, {Figure 1 -  Cumulative apparent diameter distribution of
galaxies in the ESGC. The dashed line represents the expected
distribution under the assumption that galaxies are uniformly
distributed in space.}

\fig esgc_diam_compare.ps, 6, {Figure 2 -  {\it a)} Correlation between the
original ESO
diameters and the ESGC diameters for the 347 objects in common between
the two datasets. {\it b)} As in {\it a)} for the  595 galaxies in
common between UGC and ESGC. }

\fig eso_diam_compare.ps, 6, {Figure 3 -  The relation between the original ESO
diameters,
and those corresponding to the ${\rm 26^{th}}$ isophotal level in
B from ESO-LV. All ESO-LV galaxies for which the latter is available are
plotted, for a total of 14864 objects.}

\fig n_of_z.ps, 6, {Figure 4 -  Redshift distributions for the different ORS
subsamples. The distribution for the \iras\ 1.2 Jy survey is also
shown for comparison.  The number of galaxies in each subsample has
been normalized by their respective solid angles, and is shown in
logarithmic bins of width 0.1 dex.}

\fig exclude_unobserved.ps, 6, {Figure 5 -  a. Regions not covered by ORS are
shown in an Aitoff
projection on the sky using Galactic coordinates. The horizontal strip
corresponds to the zone of avoidance ($|b| \leq 20^\circ$). The heavy
lines mark the boundaries of the ESGC strip. The dark areas correspond
to high extinction regions ($A_B > 0.7$ mag) and the grey ones to
those plates in the Southern sky not surveyed by the ESO-LV, and
therefore not covered by ESO-m. b. ORS galaxies for which no redshift is yet
available
are shown in an Aitoff projection on the sky.
Different symbols are used for galaxies belonging to ORS-m and to
ORS-d. Most of the missing redshifts are in ORS-d and correspond to
faint or low surface brightness galaxies. The ORS region with the largest
degree of incompleteness is the ESGC strip, which is contained within the
two heavy lines. There are 171 galaxies in all without redshifts. }

\fig ors_all_temp.ps, 4, {Figure 6a -  The ORS sample shown
in polar projections of the two Galactic hemispheres. The circle
on the left (right)
correspond to the North (South) Galactic Hemisphere, as
indicated. Galactic latitude decreases radially from the center of
each panel, whereas Galactic longitude runs in azimuth. The dashed
circles are at constant Galactic latitude, and are spaced every
$20^\circ$ starting at $b = 20^\circ$ and $b = -20^\circ$.
The outer rings devoid of galaxies in both hemispheres
correspond to the zone of avoidance. The two heavy lines in each
circle demark the ESGC strip. The excluded regions are
indicated with hatching (compare Figure~5a). b. As in Figure 6a, now in Aitoff
projection. }

\fig hemipair2.ps, 6, {Figure 7a -  The upper panel is as in Figure 6a, but now
showing only those galaxies with redshifts between 0 and 3000 \kms.
The lower panel shows the galaxy density field in the redshift shell
at 1500 \kms. The heavy solid contour is at mean density; the dashed
contours are at $-1/3$ and $-2/3$ the mean density, and overdense
regions are indicated by solid contours logarithmically spaced so that
three contour levels represents an increase of a factor of two.
Several features of the density field are marked: V${}={}$Virgo
Cluster, UM${}={}$Ursa Major Cluster,
DFE${}={}$Doradus-Fornax-Eridanus complex,
TPI${}={}$Telescopium-Pavo-Indus supercluster foreground, and
LV${}={}$Local Void.}

\fig aitoffpair2.ps, 6, {Figure 7b -  As in Figure 7a, but with Aitoff
projection.}

\fig hemipair5.ps, 6, {Figure 8a - As in Figure~7a, now including galaxies
between
3000 and 6000 \kms, and showing the galaxy density field at 4500 \kms.
The structures marked are: TPI${}={}$Telescopium-Pavo-Indus supercluster, and
N${}={}$N1600; H${}={}$Hydra,
Cen${}={}$Centaurus, Ce${}={}$Cetus, Cam${}={}$Camelopardalis,
Can${}={}$Cancer, V3${}={}$Void 3 of da Costa \etal (1988).}

\fig aitoffpair5.ps, 6, {Figure 8b - As in Figure~8a, in Aitoff projection.}

\fig hemipair8.ps, 6, {Figure 9 - As in Figure~8a, now including galaxies
between
6000 and 9000 \kms, and showing the galaxy density field at 7500 \kms.
The structures marked are: Co${}={}$Coma-A1367; GW${}={}$Southern
extension of the Great Wall, S5${}={}$Supercluster 5 of Saunders \etal
(1991). }

\fig aitoffpair8.ps, 6, {Figure 9b - As in Figure~9a, in Aitoff projection.}

\fig bslice2.ps, 6, {Figure 10 -  The slice corresponding to $ -30^\circ \leq b
\leq -20^\circ$ out to 8000 \kms. Galactic longitude runs azimuthally
in the counter-clockwise direction. The Local Group lies in the
center of the plot. Redshift distances, corrected
for our motion relative to the Local Group, increase radially
outwards. The dashed circles are drawn at 2000, 4000, and 6000 \kms.
The pie-shaped wedges correspond approximately to regions of
high Galactic extinction (compare Figure 5a). }

\fig bslice1.ps, 6, {Figure 11 -  As in Figure 10, for the slice
$ 20^\circ \leq b \leq 30^\circ$.}

\fig decslice.ps, 6, {Figure 12 -  As in Figure 10, for the
ESGC strip $ -17.5^\circ \leq \delta \leq -2.5^\circ$.}

\end